# Comprehensive investigation of Quantum Oscillations in Semimetal Using an ac Composite Magnetoelectric Technique with Ultrahigh Sensitivity

Long Zhang[1,2,†], Tianyang Wang[3,†], Yugang Zhang[1,2], Shuang Liu[1], Yuping Sun[3,4], Xiaoyuan Zhou[1], Young Sun[1], Mingquan He[1,2], Aifeng Wang[1,2], Xuan Luo[3,*], Yisheng Chai[1,2,*]

[1]*Center of Quantum Materials and Devices, Chongqing University, Chongqing 401331, China.*

[2]*Low Temperature Physics Laboratory, College of Physics, Chongqing University, Chongqing 401331, China*

[3]*Key Laboratory of Materials Physics, Institute of Solid State Physics, HFIPS, Chinese Academy of Sciences, Hefei, 230031, China.*

[4]*Anhui Province Key Laboratory of Condensed Matter Physics at Extreme Conditions, High Magnetic Field Laboratory, HFIPS, Chinese Academy of Sciences, Hefei, 230031, China.*

\*Corresponding email: xluo@issp.ac.cn and yschai@cqu.edu.cn,

[†]These authors contributed equally to this work

## Abstract

Quantum oscillation (QO), a physical phenomenon that reflects the characteristics of the Fermi surface and transport fermions, has been extensively observed in metals and semimetals through various approaches, like magnetostriction, magnetization, resistivity, and thermoelectric power. However, only some allowed oscillation frequencies can be revealed by each individual method, particularly in semimetals with intricate Fermi pockets and associated magnetic breakdown phenomena. In this paper, we present the application of an ac composite magnetoelectric (ME) technique to measure the QOs of a topological nodal-line semimetal, ZrSiS, which possesses six fundamental QO frequencies. By employing the ME technique with a maximum magnetic field of 13 T and a minimum temperature of 2 K, we are able to capture all the fundamental frequencies and most of the permissible magnetic breakdown frequencies. In comparison, some of the frequencies were missing in the aforementioned four methods under identical measurement conditions. Remarkably, a series of magnetic breakdown frequencies around 8000 T were revealed even in a magnetic field as low as 7.5 T. These findings highlight the ME technique as an ultrahigh-sensitive tool for studying Dirac Fermions and other topological semimetals with complex Fermi surfaces.

# Introduction

In metals or semimetals, application of a strong magnetic field ($B$) at low temperatures ($T$) leads to the quantization of electronic states into discrete Landau Levels with distinct energies, resulting in periodic oscillations of the density of states near the Fermi surface, dubbed Quantum Oscillation (QO)[1]. These oscillations were first experimentally observed in magnetization and resistivity measurements in bismuth during the 1930s, known as de Haas–van Alphen (dHvA) oscillations and Shubnikov–de Haas (SdH) oscillations, respectively[2,3]. The oscillatory period is traditionally represented in an inverse magnetic field plot. The corresponding frequency ($F_r$) is proportional to the extremal cross-section of the Fermi surface ($A_F$) perpendicular to the applied field, as dictated by the Onsager relation $F_r = (\varphi_0/2\pi^2)A_F$, where $\varphi_0$ is the magnetic flux quantum. This characteristic establishes Quantum Oscillation as one of the most powerful tools for investigating the geometry of Fermi surface.

In recent years, nodal-line semimetals (NLSMs) that feature Dirac band crossings as lines or closed loops in momentum space have attracted significant interest as new topological systems[4-9]. Among the experimentally verified NLSMs, ZrSiS stands out with its strong linear Dirac dispersion spanning an energy range of ~2 eV, indicating the presence of nearly massless transport fermions[6,7]. It is also notable for its intricate Fermi surface configuration and up to six fundamental frequencies: $F(\delta_1)$ ~ 8.5 T, $F(\delta_2)$ ~ 16.5 T, $F(\gamma_1)$ ~ 22 T, $F(\gamma_2)$ ~ 92 T, $F(\alpha)$ ~ 240 T, $F(\beta)$ ~ 419 T[16]. Accordingly, clear QOs have been detected in various physical properties, including resistivity[10-12], magnetization[13], thermopower[14], and magnetic torque[15,16]. Notably, the $\alpha$ and $\beta$ pockets located on the Z-R-A plane can additionally exhibit the magnetic breakdown phenomena, leading to QO frequencies up to ~ 8000 T. However, no single method has been successful in fully capturing all these frequencies at temperatures above 2 K and in magnetic fields below 14 T. Discrepancies among different studies have only provided fragmented insights into the complete picture of the Fermi surface.

Typically, the magnitude of QOs at a specific frequency increase with declining temperatures and increasing magnetic fields. This can be ascribed to the thermal damping factor $R_T = aT\mu/[B\sinh(aT\mu/B)]$ and the field-related damping factor $R_D = \exp(-aT_D\mu/B)$ in the Lifshitz-Kosevich (LK) formula, respectively[1]. Here, $a = (2\pi^2 k_B m_0)/(\hbar e)$ is a constant, $\mu = m^*/m_0$ represents the relative effective mass, $m_0$ denotes the mass of a free electron, and $T_D$ is the Dingle temperature. Consequently, to extract comprehensive information about the Fermi surface, investigations of QOs often require ultralow temperatures (<1 K) and high magnetic fields (>20 T). However, achieving these conditions is challenging with standard cryogenic magnets, which typically operate at relatively moderate temperatures (>1.8 K) and within a limited field range (<14 T). Under these conditions, each measurement technique may only reveal a subset of QO frequencies, like in ZrSiS[11-14]. So far, a single measurement capturing all QO frequencies under moderate conditions does not exist, and becomes highly desirable for investigating the non-trivial and complex Fermiology in semimetals.

Recently, an ac composite magnetoelectric (ME) technique has emerged as a highly sensitive tool for investigating magnetic materials[17,18]. In this method, a thin sample is affixed to a piezoelectric transducer to form a composite magnetoelectric laminate. This configuration facilitates the conversion of the in-plane magnetostriction of the specimen into an electrical signal across the piezo-layer. The magnetostriction ($\lambda$) measures the response of sample geometry in external magnetic fields, i.e., $\lambda(B) = \Delta L(B)/L_0$, where $L_0$ is the sample length in zero magnetic field, and $\Delta L(B) = L(B) - L_0$ is the length change induced by magnetic field. Unlike traditional dc measurements, the ac ME method employs the ac-lock-in technique, which achieves an enhanced signal-to-noise ratio and high resolution by applying an ac-driven magnetic field ($B_{ac}$). This technique has demonstrated unprecedented accuracy in characterizing magnetic skyrmion phases in materials such as MnSi[17] and $Co_7Zn_8Mn_5$[18], and in unveiling ferromagnetic spin-spin correlations in quasi-2D magnets, even within paramagnetic phases[19]. Hence, the ac ME method has potential for detecting weak Quantum Oscillations (QOs) in the magnetostriction of semimetals under moderate field and temperature conditions.

Theoretically, the ac ME technique can achieve superior resolution in detecting QOs of magnetostriction in semimetals with high frequencies. On one hand, the real part of the ac voltage signal $V_x$ is expected to be proportional to the piezomagnetic coefficient $\mu_0(\partial\lambda)/\partial B$ (where $\mu_0$ is the permeability of vacuum) if the sample volume is sufficiently large[20]. On the other hand, for a thin crystal with a soft nature that is tightly clamped by the piezo-layer, the strain remains nearly constant under changing magnetic fields. As a result, $V_x$ is primarily generated by the stress ($\sigma$) transferred from the sample and is proportional to $\mu_0(\partial\sigma)/\partial B$ with constant strain. Consequently, the QO of $V_x$ with a frequency of $F_r$ can be deduced in two cases:

$$V_x \sim \frac{\partial \lambda}{\partial B} = -\frac{1}{B}\left[\tilde{\lambda}(\varphi_r) + \frac{2\pi F_r}{B}\tilde{\lambda}\left(\varphi_r + \frac{\pi}{2}\right)\right] \quad (4)$$

$$V_x \sim \frac{\partial \sigma}{\partial B} = -\frac{1}{B}\frac{\partial \sigma}{\partial \lambda}\left[\tilde{\lambda}(\varphi_r) + \frac{2\pi F_r}{B}\tilde{\lambda}\left(\varphi_r + \frac{\pi}{2}\right)\right] \quad (5)$$

Here, the above relations contain two terms with different amplitudes and oscillatory phases. The first term exhibits the same oscillatory frequency and phase ($\varphi_r$) as the QO of magnetostriction, while the second term displays a phase shift of $\pi/2$ with an amplification factor of $(2\pi F_r)/B$. This factor can significantly enhance the ME signal for most QO frequencies above 10 T, making the second term dominant. Notably, for high frequencies above 100 T, the onset field for observing QOs may be significantly reduced since the amplification factor becomes even larger in lower magnetic fields.

In this study, we conducted an investigation on the nodal-line semimetal ZrSiS. We compared the ac composite ME method with four other techniques, including in-plane resistivity ($\rho$), out-of-plane magnetization ($M$), out-of-plane magnetostriction ($\lambda$), and in-plane thermopower ($S$), using the same thin ZrSiS single crystal. The measurements were carried out at temperatures down to 2 K and in magnetic fields up to 13 T. Our findings demonstrate that the ME method can effectively: 1) reveal all six fundamental frequencies, which cannot be achieved by other methods; 2) display a series of magnetic breakdown frequencies around 8000 T with a low onset magnetic field of 7.6 T; and 3) accurately determine the effective mass values of $\delta_1$ and $\alpha$ bands from temperature-dependent data, showing nearly zero values. These results highlight the potential of our ME technique for exploring Dirac and Weyl physics, as well as Fermi surface properties in other metals, under moderate experimental conditions.

## Results

### Basic properties of ZrSiS

ZrSiS belongs to the WHM (W = Zr, Hf; H = Si, Ge, Sn, Sb; M = S, Se, Te) material system and crystallizes in the PbFCl structure type with a tetragonal *P*4/*nmm* space group (no. 129)[21], as depicted in Fig. 1a. The Si square net is situated in the *ab*-plane and is enclosed between Zr-S layers. X-ray diffraction analysis of a single crystal (inset of Fig. 1b) confirms the high quality of the ZrSiS samples. The diffraction pattern in Fig. 1b exhibits well-defined sharp (00L) peaks, and the calculated *c*-axis lattice constant is 8.060 Å using the Nelson-Riley method, consistent with previous reports[22,23].

Figure 1c illustrates the temperature dependence of resistivity ($\rho$) along the *a*-axis for two magnetic fields: 0 T and 8 T, where the field is aligned parallel to the *c*-axis. In the absence of a magnetic field, the resistivity shows typical metallic behavior, gradually decreasing as the temperature drops. The high quality of the crystal is evident from the calculated residual resistivity ratio of approximately 50. Conversely, in 8 T, a clear transition occurs, leading to a pronounced insulating behavior below 100 K. This field-induced crossover can be linked to the emergence of a band gap in topological semimetals, a phenomenon consistent with prior research[11].

### Quantum oscillations in magnetic properties: magnetostriction and magnetization.

As mentioned in various studies, the QOs are expected to be noticeable under high magnetic fields and low temperatures in different magnetic properties[10–16]. QOs in magnetostriction $\lambda$ is a fundamental thermodynamic phenomenon that can directly reflect strain dependence of the Fermi surface. As depicted in Fig. 2a, field-dependent magnetostriction $\lambda(B)$ for both *B* and $\lambda$ along the *c*-axis, exceeding $10^{-5}$ at 13 T, is measured with a capacitive dilatometer at 1.6 K. An oscillatory component with a minor amplitude ($10^{-7}$-$10^{-8}$) and low frequency is shown in the inset of Fig. 2a by subtracting a dominating linear background. The onset field of the oscillations is about 1.5 T. Other higher frequency oscillations reported in previous studies are not found in $\lambda(B)$ due to the thin crystal thickness (0.36 mm) and large background signal. In contrast, field-dependent magnetization *M*(*B*) (Fig. 2b) exhibits both low and high-frequency oscillations. A similar dominant linear background is also identified while the onset field of the oscillations is reduced to 1 T, indicating better sensitivity in terms of QO. As a result, these two non-transport methods yield considerably less information within limited field and temperature condition compared to the known exotic features found in QO of ZrSiS appearing in high magnetic fields[10,15,16].

**Quantum oscillations of ZrSiS in in-plane resistivity and thermopower.**

To show the QOs detected by transport properties, we measured the field-dependent resistivity $\rho$ and thermopower $S$ on the same crystal used in magnetic property measurements, as depicted in Fig. 2c and 2d, respectively. Field-dependent resistivity $\rho(B)$ (Fig. 2c) presents strong high-frequency and weak low-frequency oscillations. The dominant background of the $\rho$ curve follows a $B^2$-like profile, which is common in metals. The substantial background raises its discernable onset field of oscillations to over 4 T. Conversely, field-dependent thermopower $S(B)$ at 1.6 K reveals more complex oscillations (Fig. 2d). Its background is much smaller than that of other methods. Consequently, its oscillation part shows an onset field of the oscillations as low as 1.5 T. The 8000 T QO, which has relatively weak magnitude, appears around 8.7 T. The overall field dependent profile of $S(B)$ is consistent with previous studies[14].

**Quantum oscillations of ZrSiS with composite magnetoelectric configuration.**

We found that the composite ME method exhibits a superior sensitivity compared to the above four methods. To employ the composite ME method on ZrSiS, we mechanically bonded the single crystal with a thin $0.7Pb(Mg_{1/3}Nb_{2/3})O_3$-$0.3PbTiO_3$ (PMN-PT) [001]-cut single crystal using silver epoxy to produce a ZrSiS/PMN-PT laminate, as depicted in the inset of Fig. 2e. The real part of ME signal $V_x$ as a function of the dc magnetic field $B$ is measured at 2 K, as shown in Fig. 2e. QOs with various frequencies are displayed, revealing more details than the above four methods. $V_x$ has a weak linear background, leading to a dominant oscillation signal as the magnetic field increases. This can be understood based on Eq. (4), where $\frac{\partial \lambda}{\partial B}$ makes a linear background with a constant value. With a weak background, the onset field of the oscillations drops to 0.5 T. Additionally, $V_x$ exhibits oscillations around 8000 T, as shown in the inset of Fig. 2e, similar to that in thermopower (inset of Fig. 2d). However, a lower starting field of 7.6 T is observed in the ME technique. Based on these features, the ME technique appears to be an excellent method for characterizing the quantum oscillations.

**Feature comparison and FFT analyses of QOs in ZrSiS.**

In this study, QOs measured by different methods are quantitatively analyzed. Fast Fourier transform (FFT) analyses are employed to extract the oscillatory frequencies and amplitudes for each method, which are presented in Fig. 3k-o. The oscillations of various physical properties are plotted as a function of $1/B$ (inverse magnetic field) with the background subtracted. The entire field region is divided into two sections: high field region ($0.07 < 1/B < 0.2$ T$^{-1}$, Fig. 3a-e) and low field region ($0.2 < 1/B < 0.5$ T$^{-1}$, Fig. 3f-j). For $1/B > 0.5$ T$^{-1}$, the oscillation is too weak to be analyzed. The six fundamental frequencies are identified and marked as dashed lines in the FFT plots. The discrepancies between different methods are revealed in the comparison. Oscillation frequencies of $F(\delta_1)$ 8 T and $F(\delta_2) \sim 16$ T are common among all the methods, except in resistivity where these two frequencies are not resolvable, as shown in Fig. 3n. $F(\gamma_1) \sim 22$ T appears to exist only in the oscillations of magnetostriction and ac magneto-electric signals ($V_x$), but the FFT peak in $V_x$ is more pronounced than that in $\lambda$, with its onset field ($\sim 2.85$ T) indicated in the low field region (Fig. 3g). $F(\gamma_2) \sim 91$ T is difficult to discern in any oscillation data directly, while only the FFT result of $V_x$ (Fig. 3i) shows a discrete peak at this frequency. For the fundamental frequencies above 100 T, $F(\alpha) \sim 240$ T oscillations are common in these methods except in $\lambda$ (Fig. 3l-o). $F(\beta) \sim 420$ T only appears in $V_x$, with the smallest amplitude in the FFT profile (Fig. 3l) compared to all the other fundamental frequencies.

It is noted that, in the high-field region (Fig. 3b) of $V_x$, the oscillations exhibit a complex pattern, indicating that more frequencies are included. The FFT analysis suggests a series of oscillations in the frequency range of 60 to 600 T (65, 183, and 593

T), which are combinations between $F(\alpha)$ and $F(\beta)$ orbits due to the magnetic breakdown phenomenon[16]. Only the 593 T frequency can be found in thermopower data as well (Fig. 3o). All these fundamental and breakdown frequencies below 1 kT are summarized in Table 1 and the frequencies above 1 kT will be discussed below. From Table 1, a combination of magnetostriction, resistivity, magnetization, and thermopower is not sufficient to capture all the fundamental frequencies, and $F(\gamma_2)$ is still missing. In contrast, the ME method demonstrates an ultra-high-sensitive performance, as it independently reveals all the allowed frequencies reported previously[10-16].

## Discussion

We noticed that the effective mass $m^*$ can be estimated by analyzing the temperature-dependent amplitude of ME QOs in FFT according to the Lifshitz-Kosevich formula and Eq. 4. Therefore, the field-dependent ME signals at selected temperatures are measured and depicted in Fig. 4a. A clear damping effect of oscillation amplitudes is revealed when the temperature increases from 2 to 20 K. The corresponding FFT amplitudes of the featured frequencies also reflect this tendency, as shown in Fig. 4b. $F(\delta_1)$ and $F(\alpha)$ are chosen to fit the thermal damping factor $R_T$ in the conventional LK formula, as shown in Fig. 4c. Their temperature-dependent behaviors exhibit metallic fermion, and the effective masses are extracted to be $m^*(\delta_1) = 0.0357$ $m_0$ and $m^*(\alpha) = 0.0451$ $m_0$ ($m_0$ is the mass of a free electron). Such small effective masses indicate nearly massless fermions for these two orbits, consistent with previous de Haas-van Alphen oscillation studies[13].

A series of ultra-high frequencies can also be observed in ME signals (Figure 4d), which have been reported by magnetic torque and resistivity up to 35 T and down to 0.34 K[10,15,16]. The most pronounced peaks appear in a region of around 8.5 kT, labeled with six Roman numbers I to VI. The differences in frequency between these distinct peaks are ~ $F(\alpha)$ except the one between peaks IV and V, which is ~ $F(\beta) - F(\alpha)$. Two more peaks in a frequency region of around 11 kT are also detected but with much reduced amplitudes, and the difference between them is ~ $F(\alpha)$. The gap in frequency between these two regions is indicated with a frequency ~ 2330 T, whose origin is still controversial in the literatures[10,15,16]. These apparent features are overall consistent with the latest dHvA oscillation measurements[16], which further validate the high sensitivity of our method under moderate field and temperature conditions. The effective masses of peak III and IV are fitted to be $m^*$(III) = 0.124 $m_0$ and $m^*$(IV) = 0.159 $m_0$.

In summary, we have demonstrated that the ac composite ME method possesses high sensitivity for studying the complex QOs in the Dirac semimetal ZrSiS. In comparison to four other typical methods with the magnetic field and temperature limited to below 13 T and above 2 K, respectively, ME QOs provide the most complete configuration of Fermi surfaces in ZrSiS, whereas the combination of all four methods still lacks some of the allowed orbitals. In particular, FFT result of ME oscillations yields a series of clearly separated peaks above 7.5 kT. Meanwhile, by studying the temperature damping effect of ME QOs, the effective masses of some orbitals can be obtained. These investigations suggest that the ac composite ME method is an ultrahigh-sensitive and powerful tool for studying Dirac Fermions and other topological semimetals under moderate magnetic field and temperature conditions.

## Methods

### Sample preparation.

Single crystals of ZrSiS were synthesized by the chemical vapor transport method. The stoichiometric mixture of Zr (Alfa Aesar, 99.9 %), Si(Alfa Aesar, 99.9 %) and S(Alfa Aesar, 99.9 %) powders were sealed in an evacuated quartz tube with $I_2$ being used as transport agent. The sealed quartz tubes were put in a two-zone tube furnace. The hot side is about 1020 °C and the cold side is 920 °C, and dwelled for two weeks. Plate-like shape single crystals with shinning surfaces were obtained. The size of the crystal was about 2*3*0.5 mm$^3$.

### Transport measurements.

Temperature dependent resistivity was measured using a four-probe method. All the measurements were done in a Quantum Design Physical Properties Measurement System for (PPMS-9T) 1.8 K<$T$<400 K and $H$<9 T.

### Structural characterization.

X-ray diffraction measurement of single crystal was performed by the PANalytical X'Pert diffractometer using the Cu $K_α$ radiation (λ = 0.15406 nm) at room temperature.

### Magnetization measurements.
Magnetization measurements were conducted on oriented crystals in steady magnetic fields up to 14 T and at a temperature of 2 K. Either a commercial superconducting quantum interference device (SQUID) or a vibrating-sample-magnetometry from Quantum Design is used.

### Striction measurements.
High resolution magnetostriction measurements were achieved by using a capacitive dilatometer made of CuBe. To precisely measure the capacitance change, an Andeen Hagerling 2550A ($f$ = 1 kHz) capacitance bridge with driven voltage of 15 V was used.

### Composite ME measurements.
The configuration of this method is shown in Fig. 2b. The sample was glued onto a piezoelectric transducer (PMN-PT) with silver epoxy. In-plane magnetostriction of the sample can be converted into an electrical signal across the piezo-layer. A constant ac magnetic field Hac (~ 0.5 Oe, $f$ = 211 Hz) was applied through a Helmholtz coil along with the static field H. The induced piezoelectrical ac signal was measured with a lock-in amplifier (OE1022, SYSU Scientific Instruments).

**Thermoelectric power measurements.** The thermoelectric power ($S$) was measured along the *a*-axis of a single crystal of dimensions 2.72 × 2.4 × 0.36 mm$^3$(S2) with the magnetic field (B) parallel to the *c*-axis. The sample is bridged over two oxygen-free copper blocks to generate a temperature difference. When measuring the quantum oscillations, heat power was held constant to maintain a temperature difference of ~ 0.5 K for the sample.

**FFT analysis.** These analyses were performed using the rectangle window function to prevent the modification of oscillatory amplitude. Appropriate field regions were chosen to ensure signal-to-noise ratio, with suitable zero padding to ensure resolution of the X-axis (frequency) of the results.

## Acknowledgements

This work was supported by the Natural Science Foundation of China under Grants No. 11974065, No. 12227806, and No. 11874357, the National Key Research and Development Program under Contract No. 2021YFA1600201, the Joint Funds of the National Natural Science Foundation of China and the Chinese Academy of Sciences' Large-Scale Scientific Facility under Contracts No. U1932217. Y.S.C. acknowledges the support from Beijing National Laboratory for Condensed Matter Physics. We would like to thank G. W. Wang and Y. Liu at Analytical and Testing Center of Chongqing

**Figure and Figure captions**

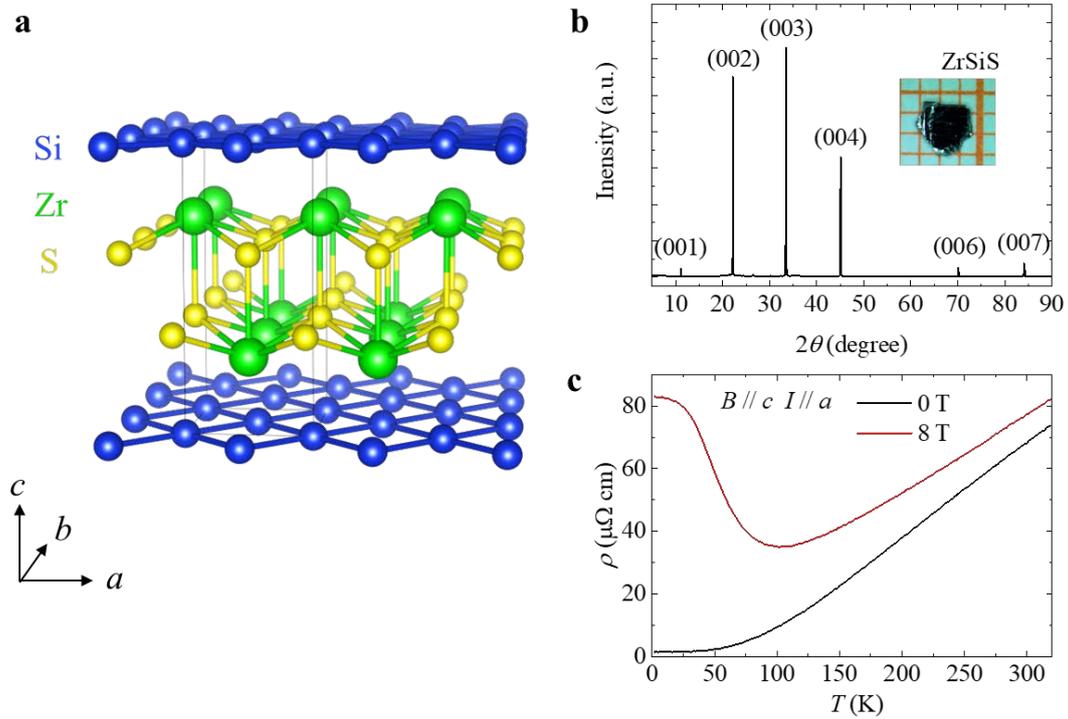

**Figure 1 | Structure and basic sample characterizations of ZrSiS single crystal.** (**a**) Schematic crystal structure of ZrSiS with P4/*nmm* space group. (**b**) Single crystal X-ray diffraction pattern measured at room temperature. Inset: A ZrSiS single crystal. (**c**) Temperature-dependent resistivity $\rho$ under external magnetic fields of 0 and 8 T along *c*-axis.

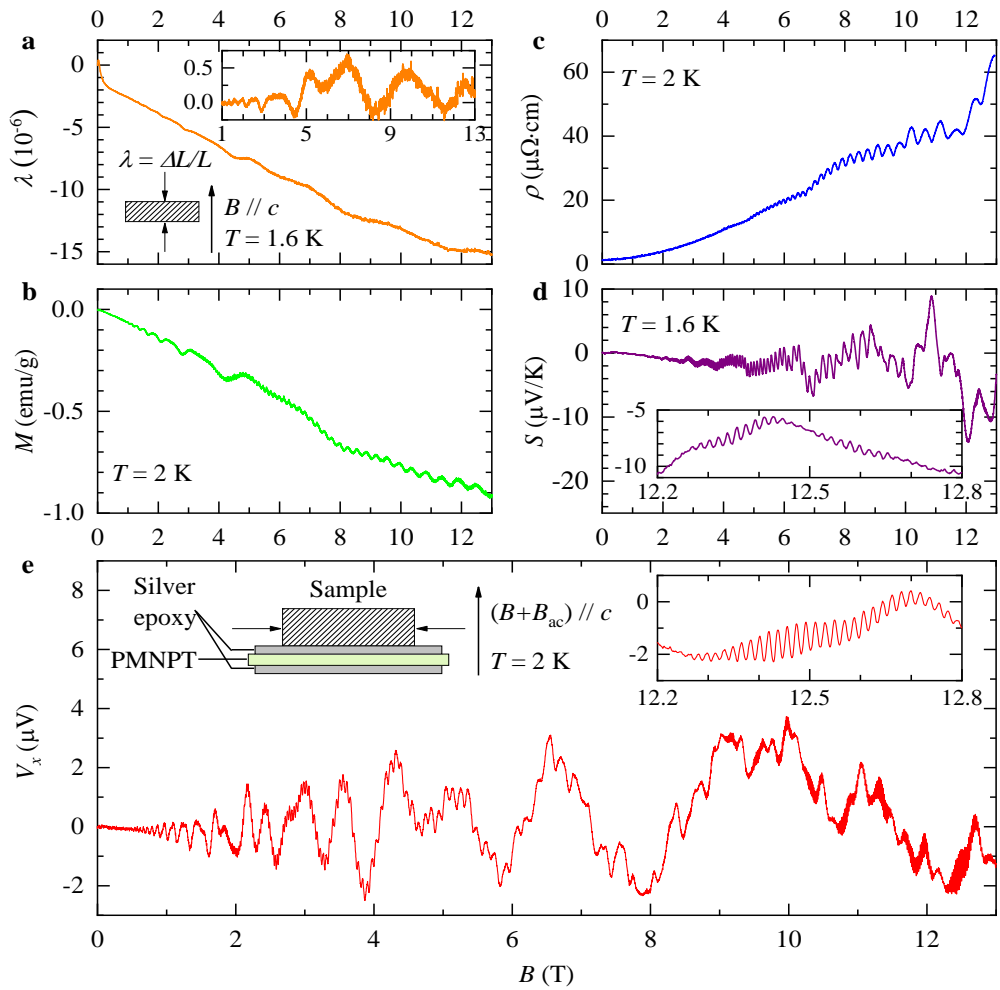

**Figure 2 | The magnetic field dependences of five physical properties in ZrSiS at around 2 K with the magnetic field parallel to the *c*-axis.** (**a**) Field-dependent magnetostriction along the *c*-axis ($L = 0.36$ mm). Inset of (a) displays the oscillations part with the background subtracted. (**b**) Field-dependent magnetization. (**c**) Field-dependent resistivity. (**d**) Field-dependent thermopower. Inset of (d) also shows an ultra-high frequency oscillation in the field range above 12 T. (**e**) Real part of ac voltage signal stimulated from in-plane deformation of sample by gluing the sample to PMNPT. The deformation is induced by a sinusoidal magnetic field $B_{ac}$ (left panel of inset of (e)). Right panel of inset of (e) shows an ultra-high frequency oscillation in the field range above 12 T.

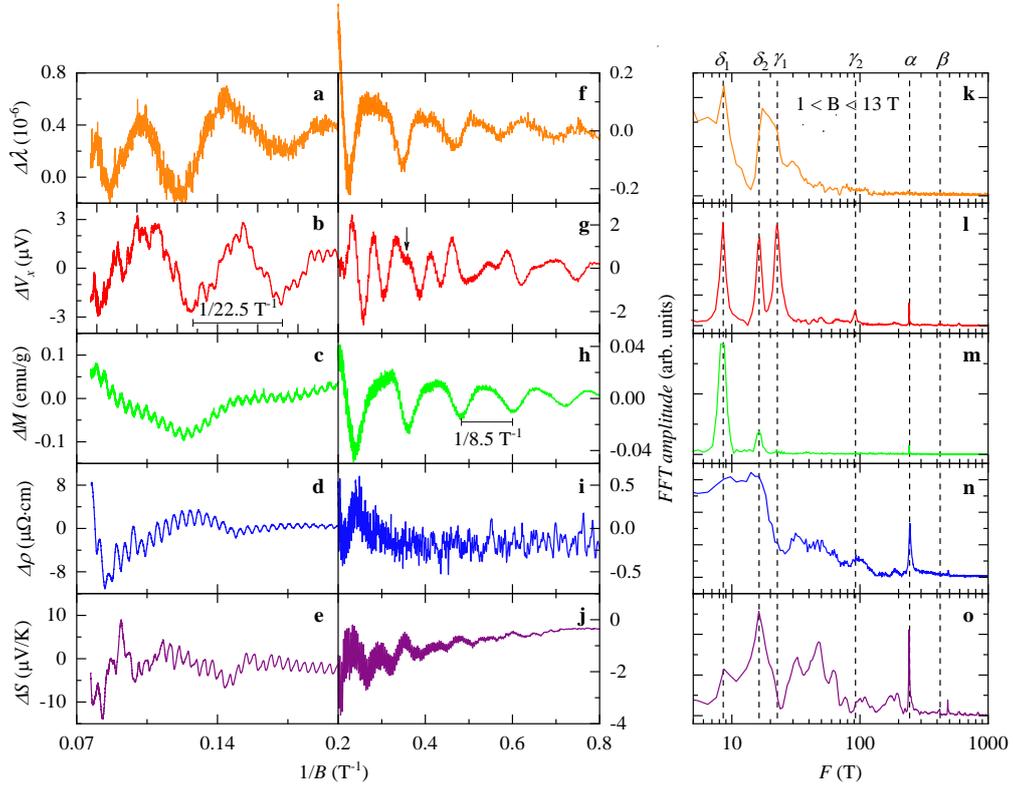

**Figure 3 | 1/B plots and FFT analyses of the oscillations of five physical properties in ZrSiS.** (**a-e**) 1/B values from 0.07 T$^{-1}$ to 0.25 T$^{-1}$. (**f-j**) 1/B values from 0.25 T$^{-1}$ to 0.8 T$^{-1}$. (**k-o**) FFT analyses of five physical properties. The six dotted lines indicate the six fundamental oscillatory frequencies in ZrSiS.

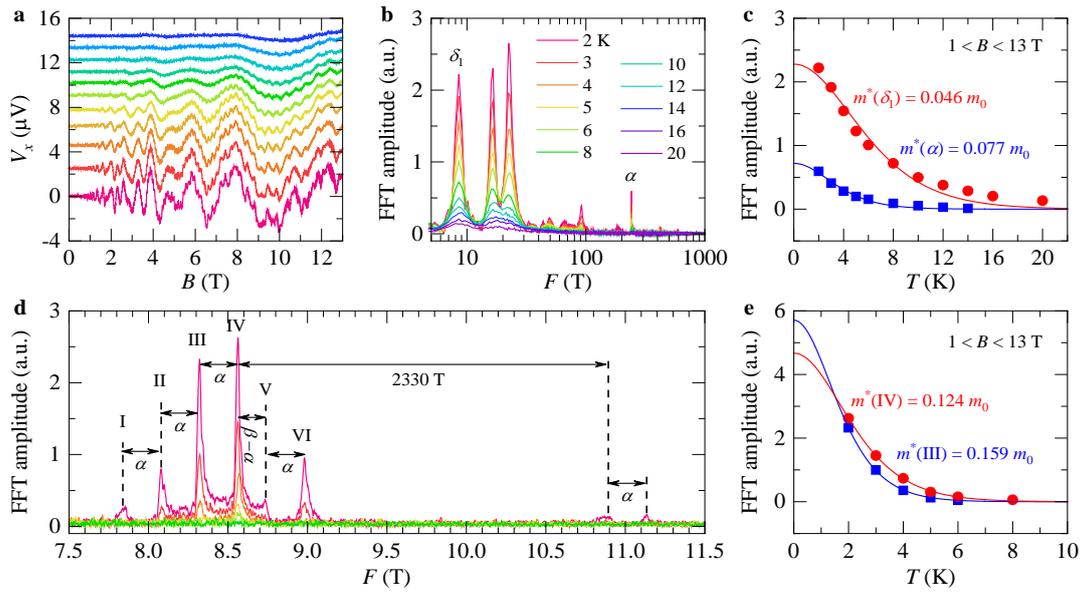

**Figure 4 | Field dependent oscillations of $V_x$ at selected temperatures and related FFT analyses.** (**a**) Background subtracted field dependent oscillations of $V_x$ at selected temperatures. (**b**) FFT spectrum (5 - 1000 T) in the magnetic field range of 1 - 13 T. (**c**) Effective mass fitting of $\delta_1$ and $\alpha$ obitals, revealing a conventional metallic Lifshitz-Kosevich behavior. (**d**) FFT spectrum (7.5 – 11.5 kT) in magnetic field range of 1 - 13 T. (**e**) Effective mass fitting of III and IV orbitals.

# Tables

**Table 1 | Main oscillatory frequencies in FFT analyses for ZrSiS**

| Method | Fundamental orbits | | | | | | Breakdown orbits (< 1000 T) | | |
|---|---|---|---|---|---|---|---|---|---|
| | $\delta_1$ | $\delta_2$ | $\gamma_1$ | $\gamma_2$ | $\alpha$ | $\beta$ | $2\alpha - \beta$ | $\beta - \alpha$ | $2\beta - \alpha$ |
| $\lambda$ | 8.7 | 17.3 | 21.7 | | | | | | |
| $V_x$ | 8.5 | 16.5 | 22.4 | 91.6 | 242.0 | 419 | 65 | 183 | 593 |
| $M$ | 8.5 | 16.4 | | | 241.1 | | | | |
| $\rho$ | | 16.3 | | | 245.0 | | | | |
| $S$ | 8.7 | 16.3 | | | 241.6 | 419 | | | 594 |

# Supplementary Material: Comprehensive investigation of Quantum Oscillations in Semimetal Using an ac Composite Magnetoelectric Technique with Ultrahigh Sensitivity

## Deduction of ac magnetoelectric voltage coefficients

In conventional band theory, when taking account of the elastic properties in the phenomenon of QOs, we can express the thermodynamic potential $\Omega$ (considering oscillations with any single oscillatory frequency $F_r$) as a function of the stress tensor $\sigma$ (strain $\lambda$):

$$\Omega = \widetilde{\Omega} + \Omega_0 \equiv \Omega_r \cos\left(\frac{2\pi F_r}{B} + \varphi_r\right) - \frac{1}{2}s\sigma^2 \quad \text{(for given stress)}$$

$$\Omega = \widetilde{\Omega} + \Omega_0 \equiv \Omega_r \cos\left(\frac{2\pi F_r}{B} + \varphi_r\right) - \frac{1}{2s}\lambda^2 \quad \text{(for given strain)} \quad (1)$$

where $s$ is elastic compliance of the sample which is considered to be a constant, $\varphi_r$ is the corresponding oscillatory phase, $\Omega$ is composed by the oscillatory $\widetilde{\Omega}$ and static $\Omega_0$ thermodynamic potential respectively. The strain $\lambda$ for given stress $\sigma$ and the stress $\sigma$ for given strain $\lambda$ are then given by:

$$\lambda = -\frac{\partial \Omega}{\partial \sigma} = \widetilde{\lambda} + s\sigma$$

$$\sigma = -\frac{\partial \Omega}{\partial \lambda} = \widetilde{\sigma} + \frac{\lambda}{s} \quad (2)$$

respectively, where

$$\widetilde{\lambda}(\varphi_r) = -\frac{\partial \widetilde{\Omega}}{\partial \sigma} = \Omega_r \frac{2\pi}{B} \frac{\partial F_r}{\partial \sigma} \sin\left(\frac{2\pi F_r}{B} + \varphi_r\right)$$

$$\widetilde{\sigma}(\varphi_r) = -\frac{\partial \widetilde{\Omega}}{\partial \lambda} = \Omega_r \frac{2\pi}{B} \frac{\partial F_r}{\partial \lambda} \sin\left(\frac{2\pi F_r}{B} + \varphi_r\right) \quad (3)$$

Finally, the resultant ME signal $V_x$ would be proportional to the partial differentiation of $\sigma$ with respect to field $B$, with the assumption which is reasonable for large quantum numbers that $\Omega_r$ can be treated as a constant during differentiation:

$$V_x \sim \frac{\partial \lambda}{\partial B} = -\Omega_r \left[\frac{2\pi}{B^2} \frac{\partial F_r}{\partial \sigma} \sin\left(\frac{2\pi F_r}{B} + \varphi_r\right) + \frac{4\pi^2 F_r}{B^3} \frac{\partial F_r}{\partial \sigma} \cos\left(\frac{2\pi F_r}{B} + \varphi_r\right)\right]$$

$$= -\frac{1}{B}\left[\tilde{\lambda}(\varphi_r) + \frac{2\pi F_r}{B}\tilde{\lambda}\left(\varphi_r + \frac{\pi}{2}\right)\right] \quad (4)$$

$$V_x \sim \frac{\partial \sigma}{\partial B} = -\Omega_r\left[\frac{2\pi}{B^2}\frac{\partial F_r}{\partial \lambda}\sin\left(\frac{2\pi F_r}{B} + \varphi_r\right) + \frac{4\pi^2 F_r}{B^3}\frac{\partial F_r}{\partial \lambda}\cos\left(\frac{2\pi F_r}{B} + \varphi_r\right)\right]$$

$$= -\frac{1}{B}\frac{\partial \sigma}{\partial \lambda}\left[\tilde{\lambda}(\varphi_r) + \frac{2\pi F_r}{B}\tilde{\lambda}\left(\varphi_r + \frac{\pi}{2}\right)\right] \quad (5)$$

where $\frac{\partial \sigma}{\partial \lambda}$ can be regarded as a constant up to the first order, the first term will have the same oscillatory frequency and phase as those of the magnetostriction measurement. The second term exhibits a $\pi/2$ phase shift with a complex amplitude tuning by magnetic field $B$ and oscillatory frequency $F_r$. Therefore, we expect to observe more oscillations in $V_x$ than in $\lambda(B)$ since this method is highly sensitive.